\documentclass[10pt]{article}

\usepackage{cite,epsfig}
\def\be{\begin{equation}}
\def\ee{\end{equation}}
\def\bea{\begin{eqnarray}}
\def\eea{\end{eqnarray}}
\def\nn{\nonumber}

\newcommand{\idel}{i\,\delta}
\newcommand{\eps}{\epsilon}
\newcommand{\rd}{{\mathrm{d}}}
\newcommand{\loopint}[1]{\int \!\!\! \frac{d^D #1}{\left(2\pi\right)^D}\!}

\begin{document}

\begin{titlepage} 
\vspace*{-1cm} 
\begin{flushright} 
IPPP/08/16\\
DCPT/08/32\\
\end{flushright} 
\vskip 3.5cm 

\begin{center} 
{\Large\bf Sector Decomposition}

\vskip 1.cm 

{\large Gudrun Heinrich}

\vskip 0.7cm

{\it IPPP, Department of Physics, 
University of Durham,\\ South Road,
Durham DH1 3LE, UK\\[1mm]
gudrun.heinrich@durham.ac.uk}

\end{center} 
\vskip 4cm 

\begin{abstract}
Sector decomposition is a constructive method to isolate 
divergences from parameter integrals 
occurring in perturbative quantum field theory. 
We explain the general algorithm in detail and
review its application to multi-loop Feynman parameter
integrals as well as infrared divergent phase-space integrals
over real radiation matrix elements.
\end{abstract}


\end{titlepage} 

\newpage 

\section{Introduction}\label{intro}

Modern particle physics has reached a level of experimental accuracy
in the percent range, and some present and future precision 
experiments require theoretical uncertainties to be 
at the permille level. 
This need for precise theory predictions has pushed forward 
the frontier for calculations of higher orders in
perturbation theory  considerably in recent years. 
The calculation of higher order corrections 
relies to a large extent on tree-- or loop level 
Feynman diagrams, where  the unobserved degrees of freedom, respectively the 
loop momenta,   have to be integrated out.
It is well known that these integrations become 
increasingly difficult at higher orders, 
as the dimensionality 
of the integration parameter space and/or the number of scales  
involved is growing. 
The intricacy  
is closely related to the fact that these integrals in general contain
ultraviolet (UV) or infrared\footnote{We will use ``infrared" 
to denote both soft and collinear divergences.} (IR) divergences
which need to be renormalised respectively factorised, 
because they hinder an immediate numerical evaluation 
of complicated expressions.
The subtraction of these singularities becomes 
more and more cumbersome at higher orders,
due to the fact that the divergences will be entangled 
in an increasingly complicated way.

For ultraviolet divergences, Bogoliubov, Parasiuk, Hepp and 
Zimmermann~\cite{Bogoliubov:1957gp,Hepp:1966eg,Zimmermann:1969jj},  
established a subtraction 
scheme valid to all orders in perturbation theory. 
In fact, the original idea of sector decomposition 
goes back to the proof of the BPHZ theorem
by  Hepp~\cite{Hepp:1966eg}, who used a decomposition 
of integration parameter space into certain sectors in order to 
disentangle overlapping ultraviolet singularities. 

Concerning infrared divergences, the finiteness of 
sufficiently inclusive observables is guaranteed by the 
Kinoshita-Lee-Nauenberg theorem~\cite{Kinoshita:1962ur,Lee:1964is}, 
but in most practical applications, 
more exclusive information about the final state is required, 
such that a subtraction scheme for IR divergences has to be 
established. 
A  constructive scheme to do so to all orders in perturbation theory 
has not been formulated in full generality yet, 
at least in what concerns soft {\it and collinear} divergences.
However, the sector decomposition algorithm as outlined below 
offers a constructive method to isolate these divergences
from Feynman parameter integrals, which in principle is valid to all orders
in perturbation theory.

For the subtraction of (soft) IR divergences,  
several approaches have been suggested. 
Early ones, 
which are based already on the subdivision 
of the integration 
parameter space into different sectors where 
the parameters go to zero in an ordered way,
can be found e.g. in \cite{Pohlmeyer:1974ph,Breitenlohner:1975hg,Breitenlohner:1976te,Speer:1975dc,Speer:1977uf}. 
These sectors,
associated with certain sets of  one-particle irreducible subgraphs,
 are more advanced than the sectors needed in the UV case, 
and provide a resolution of the singularities without the need for an
 iterative procedure for diagrams with off-shell external momenta.
Further, the so-called 
$R^*$-operation~\cite{Chetyrkin:1982nn,Chetyrkin:1984xa} has been developed, 
which removes not only UV-divergences 
but also soft IR divergences by a procedure similar to the 
$R$-operation~\cite{Bogoliubov:1957gp} in the UV case.
For a review of these methods, see e.g. \cite{Smirnov:1991jn,Smirnov:2004ym}.

Later, the decomposition into sectors  
has been employed to 
extract logarithmic mass singularities from massive multi-scale integrals 
in the high energy limit at two loops~\cite{Roth:1996pd}. 
In~\cite{Binoth:2000ps}, the concept of sector decomposition 
has been elaborated 
to a general algorithm in the context of dimensional regularisation, 
allowing to isolate ultraviolet as well as infrared singularities 
from parameter integrals in an automated way~\cite{public_code}. 
The algorithm also has been implemented in a public code 
available from~\cite{Bogner:2007cr}.

At next-to-leading order (NLO) in perturbation theory, 
the use of dimensional regularisation~\cite{'tHooft:1972fi,Bollini:1972ui} 
and the universal infrared structure of 
gauge theories involving massless particles,  
like QED and QCD, allowed to 
establish a framework to isolate IR singularities analytically, 
leading to a multitude of phenomenological 
predictions.  
More recently, an increasing number of results 
at  next-to-next-to-leading order (NNLO) has also become available, 
and quite a number of them made use of the sector decomposition technique
to isolate infrared singularities
or to check analytical
results~\cite{GehrmannDeRidder:2003bm,Anastasiou:2003gr,Binoth:2004jv,Anastasiou:2004qd,Denner:2004iz,Anastasiou:2004xq,Czakon:2004wm,Anastasiou:2005qj,Anastasiou:2005pn,Heinrich:2006sw,Melnikov:2006di,Anastasiou:2006hc,Czakon:2006pa,Awramik:2006uz,Awramik:2006ar,Boughezal:2006xk,Denner:2006jr,Melnikov:2006kv,Anastasiou:2007mz,Anastasiou:2008ik,Melnikov:2008qs,Anastasiou:2008rm}. 

As most problems encountered in the calculation 
of perturbative higher order corrections can be reduced to the evaluation 
of multi-dimensional parameter integrals, the applicability of sector 
decomposition is quite universal.

\subsection{Multi-loop diagrams}

The first application of the algorithm presented in \cite{Binoth:2000ps}
was the numerical evaluation of massless 
two-loop  box diagrams at certain Euclidean points\footnote{By ``Euclidean" 
we mean that all kinematic invariants formed from external momenta 
are negative.}. 
Pioneering the analytic calculation of two-loop box diagrams, 
the planar topology with light-like external legs 
has been calculated  by
 V.A.~Smirnov~\cite{Smirnov:1999gc}, 
the non-planar one by J.B.\,Tausk~\cite{Tausk:1999vh}, where the numerical 
results obtained by sector decomposition served as an important 
check of the analytical calculation. 
For massless two-loop box diagrams with one off-shell leg, 
the numerical results of Ref.~\cite{Binoth:2000ps} 
were predictions which played a major role to validate
the subsequent analytical calculations~\cite{Gehrmann:2000zt,Gehrmann:2001ck}, 
finally allowing the calculation of the full two-loop QCD matrix element for 
$e^+ e^- \to 3$ jets~\cite{Garland:2001tf}.

Subsequently, sector decomposition was used to  check a considerable 
number of analytical results for 
two-loop~\cite{Smirnov:2000ie,Smirnov:2001cm,Davydychev:2002hy,Binoth:2003ak,Heinrich:2004iq,Czakon:2004wm,Awramik:2006ar,Awramik:2006uz}, 
three-loop~\cite{Smirnov:2003vi,Gehrmann:2006wg,Heinrich:2007at} and
four-loop~\cite{Binoth:2003ak,Boughezal:2006xk} 
diagrams\footnote{In Refs.~\cite{Czakon:2004wm,Awramik:2006ar,Awramik:2006uz,Boughezal:2006xk} the checks have been performed with 
an implementation of sector decomposition
by M.~Czakon, independent from the one
in~\cite{Binoth:2000ps}.}.

Sector decomposition also has been combined with other methods for an 
efficient numerical calculation 
of one-loop multi-leg amplitudes, first on a diagrammatic level in
 Refs.~\cite{Ferroglia:2002mz,Binoth:2002xh}, 
later for whole amplitudes in  Refs.~\cite{Lazopoulos:2007bv,Lazopoulos:2007ix}. 
The latter approach contains a combination of sector decomposition and contour 
deformation~\cite{Soper:1998ye,Soper:1999xk,Binoth:2005ff,Nagy:2006xy}, 
which allows to integrate the Feynman parameter representation of an amplitude 
numerically in the physical region.  
It also contains the application of sector decomposition to phase space integrals,  
which will be discussed below.
Similar ideas, i.e. the combination of sector decomposition and contour deformation, 
are employed in Refs.~\cite{Anastasiou:2007qb,Anastasiou:2008rm} for the numerical evaluation of 
multi-loop Feynman diagrams with infrared and   threshold   singularities.

Ref.~\cite{Denner:2004iz} describes the implementation of 
an algorithm based on  sector decomposition to extract the $1/\eps$ poles as well as the 
large logarithms of type $\ln(s/M^2)$ in the high-energy limit, allowing to 
obtain the next-to-leading logarithmic electroweak corrections of multi-loop 
diagrams.

Despite its success in practical applications, until very recently 
there was no formal proof that a  
strategy for the iterated sector decomposition can always be found such that 
the iteration is guaranteed to stop. This gap has been filled in Ref.~\cite{Bogner:2007cr}, 
by mapping the problem to Hironaka's Polyhedra game~\cite{Hironaka}.
The findings of  Ref.~\cite{Bogner:2007cr} subsequently have been used to 
prove that the coefficients of the Laurent series representing Feynman integrals 
in the Euclidean region with rational values for all invariants are 
a special class of numbers known by mathematicians as 
periods~\cite{Bogner:2007mn}.

Also on a more formal level, it is outlined in \cite{Kennedy:2007sj} that 
sector decomposition 
can be used to automate the renormalisation of 
quantum field theories, by mapping 
to the so-called Henge decomposition~\cite{Caswell:1981xt}, 
which served to provide a simple proof of the BPHZ theorem~\cite{Kennedy:1996ux}.
This mapping also allowed to establish  a direct
correspondence 
between overlapping divergences in Feynman parameter space  and in 
momentum space~\cite{Kennedy:2007sj}. 
In this context, one should also mention the formulation of renormalisation 
using Hopf algebras~\cite{EbrahimiFard:2005gx}, which provide a framework 
to describe the disentanglement of divergent subgraphs of Feynman diagrams.

\subsection{Phase space integrals}
After the results for various two-loop box diagrams had become available, the 
bottleneck to make progress in the calculation of differential NNLO cross sections 
for $1\to 3$ or $2\to 2$ processes 
was the complicated infrared singularity structure of phase space integrals
over matrix elements for the real radiation of (doubly) unresolved massless particles. 
As these phase space integrals can be written as dimensionally regularised 
parameter integrals, sector decomposition can serve to factorise entangled 
singularity structures in the case of real radiation as well. 
This idea has first been presented in~\cite{Heinrich:2002rc} and subsequently 
has been applied to 
calculate all master  four-particle phase space integrals where up to two 
particles in the final state
can become soft and/or collinear~\cite{GehrmannDeRidder:2003bm}.
Shortly after, this approach has been extended to be applicable to 
exclusive final states as well 
by expressing the functions produced by sector decomposition 
in terms of distributions~\cite{Anastasiou:2003gr}. 
A rapid development~\cite{Binoth:2004jv,Anastasiou:2004qd} 
lead to the calculation of $e^+e^-\to2$\,jets at 
${\cal O}(\alpha_s^2)$~\cite{Anastasiou:2004qd}. 
Further elaboration on this 
approach resulted in the first fully differential program to calculate an NNLO
cross section~\cite{Anastasiou:2004xq,Anastasiou:2005qj} and has 
lead to differential NNLO results for a number of 
processes meanwhile~\cite{Anastasiou:2005pn,Melnikov:2006di,Melnikov:2006kv,Melnikov:2008qs}.

\section{Basic concepts}\label{sec:basics}
\begin{figure}[htb]
\centerline{\epsfig{file=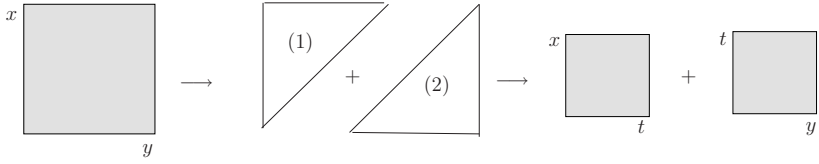,width=11.5cm}}
\caption{Sector decomposition schematically.\label{sectors}}
\end{figure}

To introduce the subject, let us look at the simple example of a two-dimensional 
parameter integral of the following form:
\begin{eqnarray}
I&=&
\int_0^1 dx\,\int_0^1dy \,x^{-1-a\epsilon}\,y^{-b\epsilon}\,\Big(x+(1-x)\,y\Big)^{-1}\;.
\end{eqnarray}
The integral contains a singular region where $x$ and $y$ vanish {\it simultaneously}, 
i.e. the singularities in $x$ and $y$ are {\it overlapping}.
Our aim is to factorise the singularities for $x\to 0$ and  $y\to 0$. 
Therefore we divide the integration range into two 
sectors where $x$ and $y$ are ordered (see Fig.~\ref{sectors})
\begin{eqnarray*}
I&=&
\int_0^1 dx\,\int_0^1dy \,x^{-1-a\epsilon}\,y^{-b\epsilon}\,\Big(x+(1-x)\,y\Big)^{-1}\,
[\underbrace{\Theta(x-y)}_{(1)}+\underbrace{\Theta(y-x)}_{(2)}]\;.
\end{eqnarray*}
Now we substitute  $y=x\,t$ in sector (1) and $x=y\,t$ in sector (2) 
to remap the integration range to the unit square and obtain
\begin{eqnarray}
I&=&\int_0^1 dx\,x^{-1-(a+b)\epsilon}\int_0^1 dt\,t^{-b\eps}
\,\Big(1+(1-x)\,t\Big)^{-1}\nn\\
&+&\int_0^1 dy
\,y^{-1-(a+b)\epsilon}\int_0^1 dt\,t^{-1-a\epsilon}\,\Big(1+(1-y)\,t\Big)^{-1}\;.
\end{eqnarray}
We observe that the singularities are now factorised such that 
they can be read off from the powers of simple monomials in the integration 
variables, while the polynomial denominator goes to a constant if the 
integration variables approach zero.
The same concept will be applied to 
$N$-dimensional parameter integrals over polynomials raised to 
some power, where the procedure in general 
has to be iterated to achieve  complete factorisation.

\section{The algorithm for multi-loop integrals}

\subsection{Feynman parameter integrals}

A general Feynman graph $G^{\mu_1\ldots\mu_R}_{l_1\ldots l_R}$ in $D$ dimensions 
at $L$ loops with  $N$ propagators  
and $R$ loop momenta in the numerator, where 
the propagators can have arbitrary, not necessarily integer powers $\nu_j$,  
has the following representation in momentum space:
\begin{eqnarray}\label{eq0}
G^{\mu_1\ldots \mu_{R}}_{l_1\ldots l_R} &=& \int\prod\limits_{l=1}^{L} \rd^D\kappa_l\;
\frac{k_{l_1}^{\mu_1}\ldots k_{l_R}^{\mu_R}}
{\prod\limits_{j=1}^{N} P_{j}^{\nu_j}(\{k\},\{p\},m_j^2)}\nn\\
\rd^D\kappa_l&=&\frac{\mu^{4-D}}{i\pi^{\frac{D}{2}}}\,\rd^D k_l\;,\;
P_j(\{k\},\{p\},m_j^2)=(q_j^2-m_j^2+i\delta)\;,
\end{eqnarray}
where the $q_j$ are linear combinations of external momenta $p_i$ and loop momenta $k_l$.
Introducing Feynman parameters according to
\bea
\frac{1}{\prod_{j=1}^N P_j^{\nu_j}}&=&
\frac{\Gamma(N_\nu)}{\prod_{j=1}^N\Gamma(\nu_j)}
\int_0^\infty \prod\limits_{j=1}^{N}\rd x_j\,\,x_j^{\nu_j-1}\,
\delta\big(1-\sum_{i=1}^N x_i \big)
\frac{1}{
\left[\sum_{j=1}^N x_j P_j\right]^{N_\nu}}\;,\\
\mbox{where }\;N_\nu&=&\sum_{j=1}^N\nu_j\;, \mbox{ leads to}\nn\\
&&\nn\\
G^{\mu_1\ldots \mu_{R}}_{l_1\ldots l_R}&=&  
\frac{\Gamma(N_\nu)}{\prod_{j=1}^{N}\Gamma(\nu_j)}
\int_0^\infty \,\prod\limits_{j=1}^{N}\rd x_j\,\,x_j^{\nu_j-1}\, 
\delta\big(1-\sum_{i=1}^N x_i\big)
\int \rd^D\kappa_1\ldots\rd^D\kappa_L\nn\\
&&k_{l_1}^{\mu_1}\ldots
k_{l_R}^{\mu_R}\,\left[ 
       \sum\limits_{i,j=1}^{L} k_i^{\rm{T}}\, M_{ij}\, k_j  - 
       2\sum\limits_{j=1}^{L} k_j^{\rm{T}}\cdot Q_j +J +\idel
                             \right]^{-N\nu}\;,\label{EQ_mixed_rep}
\eea 
where $M$ is a $L\times L$ matrix containing Feynman parameters,
$Q$ is an $L$-dimensional vector composed of external momenta and 
Feynman parameters, and $J$ contains kinematic invariants and Feynman parameters.
The  factors of $k_{l_i}^{\mu_i}$ in the numerator can be generated from $G(R=0)$
by partial differentiation with  with respect to $Q_{l}^{\mu_i}$,
where $l\in \{1,\dots ,L\}$ denotes the $l^{\rm{th}}$ component of the vector $Q$,
corresponding to the $l^{\rm{th}}$ loop momentum. 
Therefore it is convenient to define the double indices 
$\Gamma_{i}=(l,\mu_i(l))\,,\,l\in\{1,\ldots,L\}, i\in \{1,\dots ,R\}$
denoting the $i^{\rm{th}}$ Lorentz index, belonging to the $l^{\rm{th}}$ loop momentum. 

To perform the integration over
the loop momenta $k_l$, we perform the following shift in order to obtain a quadratic
form for the term in square brackets in eq.~(\ref{EQ_mixed_rep}):
\be
k_l^{\prime}=k_l-v_l\;,
\quad
v_l=\sum_{i=1}^L M^{-1}_{li} Q_i\;.
\ee
After momentum integration one obtains
\begin{eqnarray}\label{EQ:param_rep}
G^{\mu_1\ldots \mu_{R}}_{l_1\ldots l_R} &=& (-1)^{N_{\nu}}
\frac{1}{\prod_{j=1}^{N}\Gamma(\nu_j)}\int
\limits_{0}^{\infty} 
\,\prod\limits_{j=1}^{N}dx_j\,\,x_j^{\nu_j-1}\,\delta(1-\sum_{l=1}^N x_l)\nonumber\\
&&\sum\limits_{m=0}^{[R/2]}(-\frac{1}{2})^m\Gamma(N_{\nu}-m-LD/2)
\left[(\tilde M^{-1}\otimes g)^{(m)}\,\tilde l^{(R-2m)}
\right]^{\Gamma_{1},\ldots,\Gamma_{R}}\nonumber\\
&&\times\,\frac{{\cal U}^{N_{\nu}-(L+1) D/2-R}}
{{\cal F}^{N_\nu-L D/2-m}}\\
 &&\nonumber\\
\mbox{where} \qquad \quad &&\nonumber\\
{\cal F}(\vec x) &=& \det (M) 
\left[ \sum\limits_{j,l=1}^{L} Q_j \, M^{-1}_{jl}\, Q_l
-J -\idel\right]\label{DEF:F}\\
{\cal U}(\vec x) &=& \det (M) \nonumber\\
\tilde M^{-1}&=&{\cal U}M^{-1}\;,\quad
\tilde l={\cal U}\,v\nn
\end{eqnarray}   
and $[R/2]$ denotes the nearest integer less or equal to $R/2$. 
The expression \\
$[(\tilde M^{-1}\otimes g)^{(m)}\,\tilde l\,^{(R-2m)}]^
{\Gamma_{1},\ldots,\Gamma_{R}}$ stands for the sum over all different combinations 
of $R$ double-indices distributed to $m$ metric tensors and $(R-2m)$ 
vectors $\tilde l$. 
The above expression is well known~\cite{BogolShirkov,Itzykson:1980rh,nakanishi,Davydychev:1991va,Tarasov:1996bz,Smirnov:2004ym}, 
but an example to illustrate the distribution of indices may be helpful, 
so let us consider a two-loop integral where the 
$k_1$-integral is a rank two tensor 
integral and the $k_2$-integral is of rank one:
\bea
G^{\mu_1\mu_2\mu_{3}}_{112} &=& 
\int\rd^D\kappa_1\,\rd^D\kappa_2\;
\frac{k_{1}^{\mu_1}k_{1}^{\mu_2} k_{2}^{\mu_3}}
{\prod\limits_{j=1}^{N} P_{j}^{\nu_j}(\{k\},\{p\},m_j^2)}\nn\\
&=&
\frac{(-1)^{N_{\nu}}}{\prod_{j=1}^{N}\Gamma(\nu_j)}\int
\limits_{0}^{\infty} 
\,\prod\limits_{j=1}^{N}dx_j\,\,x_j^{\nu_j-1}\,\delta(1-\sum_{l=1}^N x_l)\\
&&\Big\{\Gamma(N_{\nu}-D)\;
\frac{{\cal U}^{N_{\nu}-3 D/2-3}}
{{\cal F}^{N_\nu-D}}\;\tilde l_1^{\mu_1}\tilde l_1^{\mu_2}\tilde l_2^{\mu_3}\nn\\
&&-\frac{1}{2}\,\Gamma(N_{\nu}-1-D)\;\frac{{\cal U}^{N_{\nu}-3 D/2-3}}
{{\cal F}^{N_\nu- D-1}}\,\times\nn\\
&&\left[(\tilde M^{-1}\otimes g)_{11}^{\mu_1\mu_2}\,\tilde l_2^{\mu_3}
+(\tilde M^{-1}\otimes g)_{12}^{\mu_1\mu_3}\,\tilde l_1^{\mu_2}
+
(\tilde M^{-1}\otimes g)_{12}^{\mu_2\mu_3}\,\tilde l_1^{\mu_1}
\right] \Big\}\;,\nn
\eea 
$$(\tilde M^{-1}\otimes g)^{\mu\nu}=\left(\begin{array}{cc}
\tilde M^{-1}_{11}\,g^{\mu\nu}&\tilde M^{-1}_{12}\,g^{\mu\nu}\\
\tilde M^{-1}_{21}\,g^{\mu\nu}&\tilde M^{-1}_{22}\,g^{\mu\nu}
\end{array} \right) \;.$$

The functions ${\cal U}$ and ${\cal F}$ also can be constructed
from the topology of the corresponding Feynman graph as 
follows~\cite{nakanishi,zavialov,Tarasov:1996br}. 
Cutting $L$ lines of a given connected $L$-loop graph such that it becomes a connected
tree graph $T$ defines a {\em chord} ${\cal C}(T)$ as being the set of lines 
not belonging to this tree. The Feynman parameters associated with each chord 
define a monomial of degree $L$. The set of all such trees (or {\em 1-trees}) 
is denoted by ${\cal T}_1$.  The 1-trees $T\in {\cal T}_1$ define 
${\cal U}$ as being the sum over all monomials corresponding 
to a chord ${\cal C}(T\in {\cal T}_1)$.
Cutting one more line of a 1-tree leads to two disconnected trees, or a  {\em 2-tree} $\hat T$.
${\cal T}_2$ is the set of all such  2-trees.
The corresponding chords define  monomials of degree $L+1$. Each 2-tree of a graph
corresponds to a cut defined by cutting the lines which connected the two now disconnected trees
in the original graph. The momentum flow through the lines of such a cut defines a Lorentz invariant
$s_{\hat T} = ( \sum_{j\in \rm Cut(\hat T)} p_j )^2$.   
The function ${\cal F}_0$  is the sum over all such monomials times 
minus the corresponding invariant. For a diagram with massless 
propagators, ${\cal F}={\cal F}_0$. If massive internal lines are present, 
${\cal F}$ gets an additional term as follows:    
\begin{eqnarray}\label{eq0def}	
{\cal U}(\vec x) &=& \sum\limits_{T\in {\cal T}_1} \Bigl[\prod\limits_{j\in {\cal C}(T)}x_j\Bigr]\;,\nonumber\\
{\cal F}_0(\vec x) &=& \sum\limits_{\hat T\in {\cal T}_2}\;
\Bigl[ \prod\limits_{j\in {\cal C}(\hat T)} x_j \Bigr]\, (-s_{\hat T})\;,\nonumber\\
{\cal F}(\vec x) &=&  {\cal F}_0(\vec x) + {\cal U}(\vec x) \sum\limits_{j=1}^{N} x_j m_j^2\;.
\end{eqnarray}  
${\cal U}$ is a positive semi-definite function. 
Its vanishing is related to the  UV subdivergences of the graph. 
Overall UV divergences, if present,
will always be contained in the  prefactor $\Gamma(N_{\nu}-LD/2)$. 
In the region where all invariants $s_{\hat T}$ are negative, 
which we will call the Euclidean region in the following, 
${\cal F}$ is also a positive semi-definite function 
of the Feynman parameters $x_j$.  Its vanishing does not necessarily lead to 
an IR singularity. Only if some of the invariants are zero, 
for example if some of the external momenta
are light-like, the vanishing of  ${\cal F}$  may induce an IR divergence.
Thus it depends on the {\em kinematics}
and not only on the topology (like in the UV case) 
whether a zero of ${\cal F}$ leads to a divergence or not. 
This fact makes it much harder to formulate
general theorems for the subtraction of IR singularities of  
multi-loop Feynman graphs.
The necessary (but not sufficient) conditions for an IR divergence 
are given by the Landau equations~\cite{Landau:1959fi,ELOP,Tkachov:1997ap}, 
which, in parameter space, simply mean that 
the necessary condition ${\cal F}=0$ for an IR divergence can only 
be fulfilled if some of the parameters $x_i$ go to zero, provided that 
all kinematic invariants $s_{\hat T}$ are negative.


\vspace*{3mm}

As can be seen from Eq.~(\ref{EQ:param_rep}), the difference between 
scalar and tensor integrals is, once  the Lorentz structure is extracted, 
given by the fact that there are polynomials of Feynman parameters in 
the numerator. These polynomials can simply be included into 
the sector decomposition procedure, thus treating tensor integrals
directly without reduction to scalar integrals.

However, there is yet another possibility: Any tensor integral 
can be expressed in terms of  scalar integrals in shifted 
dimensions, with some of the propagator powers different from unity, 
as has been shown in~\cite{Davydychev:1991va,Tarasov:1996br}.
As our propagators can have arbitrary 
powers $\nu_j$, and the dimension $D$ is a free parameter, 
this is a  viable alternative.

\subsection{Iterated sector decomposition}\label{itersd}

For less trivial examples than the one given in section \ref{sec:basics}, 
the singularities will not be factorised already after a single 
step of sector decomposition.
An algorithm how to iterate this procedure is described below.

Our starting point is a function of the form of Eq.~(\ref{EQ:param_rep}).
As the basic algorithm is the same for tensor integrals, we will consider 
$R=0$ here for ease of notation, i.e. 
\begin{eqnarray}\label{EQ:Gscalar}
G &=& (-1)^{N_{\nu}}
\frac{\Gamma(N_\nu-L D/2)}{\prod_{j=1}^{N}\Gamma(\nu_j)}\int
\limits_{0}^{\infty} 
\,\prod\limits_{j=1}^{N}dx_j\,\,x_j^{\nu_j-1}\,\delta(1-\sum_{l=1}^N x_l)
\,\,\frac{{\cal U}^{N_{\nu}-(L+1) D/2}}
{{\cal F}^{N_\nu-L D/2}}\;.\nn\\
\end{eqnarray}

\subsubsection*{Part I \quad Generation of primary sectors}

We split the integration domain into
$N$ parts and eliminate the $\delta$--distribution in such a way that the remaining 
integrations are from 0 to 1.  
To this end we decompose the integration range into $N$ 
sectors, where in each sector $l$, $x_l$ is largest (note that 
the remaining $x_{j\not=l}$ are not  further ordered):
\begin{eqnarray}
\int_0^{\infty}d^N x =
\sum\limits_{l=1}^{N} \int_0^{\infty}d^N x
\prod\limits_{\stackrel{j=1}{j\ne l}}^{N}\theta(x_l\ge x_j)\;.
\end{eqnarray} 
The $\theta$-function is defined as
\begin{displaymath}
\theta(x \ge y)=\left\{\begin{array}{ll}
              1& \mbox{if } x\ge y \mbox{ is true}\\
              0& \mbox{otherwise.}\end{array}
              \right.
\end{displaymath}
The integral is now split into $N$ domains corresponding   
to $N$ integrals $G_l$ from which we extract a common factor:
 $G=(-1)^{N_\nu} \Gamma(N_\nu-LD/2) \sum_{l=1}^{N} G_l$. In the  integrals $G_l$
 we substitute 
\begin{eqnarray}
x_j = \left\{ \begin{array}{lll} x_l t_j     & \mbox{for} & j<l \\
                                   x_l         & \mbox{for} & j=l \\
                                   x_l t_{j-1} & \mbox{for} & j>l \end{array}\right.
\end{eqnarray} 
and then  integrate out $x_l$ using the $\delta$--distribution.
As ${\cal U},{\cal F}$ are homogeneous of degree $L$,\,$L+1$, respectively,  
and  $x_l$ factorises completely, we have
${\cal U}(\vec x) \rightarrow {\cal U}_l(\vec t\,)\, x_l^L$ and
${\cal F}(\vec x) \rightarrow {\cal F}_l(\vec t\,)\, x_l^{L+1}$
and thus, using $\int dx_l/x_l\,\delta(1-x_l(1+\sum_{k=1}^{N-1}t_k ))=1$, we obtain 
\begin{eqnarray}\label{EQ:primary_sectors}
 G_l &=& \int\limits_{0}^{1} \prod_{j=1}^{N-1}{\rd}t_j\,t_j^{\nu_j-1}\,\,
\frac{ {\cal U}_l^{N_\nu-(L+1)D/2}(\vec{t}\,)}{ {\cal F}_l^{N_\nu-L D/2}(\vec{t}\,)} 
\quad , \quad l=1,\dots, N \;.
\end{eqnarray} 
Note that the singular behaviour leading to $1/\epsilon$--poles
still comes  from regions where a set of parameters $\{t_i\}$ 
goes to zero. This feature would be lost 
 if  the $\delta$--distribution was integrated out 
in a different way, since this would produce poles at  upper 
limits of the parameter integral as well. 
The $N$ generated sectors
will be called {\em primary} sectors in the following.
 
\subsubsection*{Part II \quad Iteration}

Starting from Eq.~(\ref{EQ:primary_sectors}) we repeat the following 
steps until a complete separation of overlapping regions is achieved.
\begin{description}
\item[II.1:] Determine a minimal set of parameters, say 
${\cal S}=\{t_{\alpha_1},\dots ,t_{\alpha_r}\}$, such that  
${\cal U}_l$, respectively  ${\cal F}_l$, vanish 
if the parameters of ${\cal S}$ are set to zero. ${\cal S}$ is 
in general not unique, and there is no general prescription which 
defines what set to choose in order to achieve a {\it minimal} number of iterations. 
Strategies to choose ${\cal S}$ such that the algorithm is guaranteed to stop 
are given in~\cite{Bogner:2007cr}. Using these strategies 
however in general leads to a
larger number of iterations than heuristic strategies to avoid infinite loops, 
described in more detail below.
\item[II.2:] Decompose the corresponding $r$-cube into $r$ {\em subsectors}
by decomposing unity according to
\begin{eqnarray}
\prod\limits_{j=1}^r \theta(1\geq t_{\alpha_j}\geq  0)=
\sum\limits_{k=1}^r \prod\limits_{\stackrel{j=1}{j\ne k}}^r 
\theta(t_{\alpha_k}\geq t_{\alpha_j}\geq 0)\;.
\end{eqnarray}
 \item[II.3:] Remap  the variables to the unit hypercube in each new 
 subsector by the substitution
 \begin{eqnarray}
t_{\alpha_j} \rightarrow 
\left\{ \begin{array}{lll} t_{\alpha_k} t_{\alpha_j} &\mbox{for}&j\not =k \\
                           t_{\alpha_k}              &\mbox{for}& j=k\,.  \end{array}\right.
\end{eqnarray}
This gives a Jacobian factor of $t_{\alpha_k}^{r-1}$. By construction
$t_{\alpha_k}$ factorises from at least one of the functions 
${\cal U}_l$, ${\cal F}_l$. The resulting subsector integrals have the 
general form
\begin{eqnarray}\label{EQ:subsec_form}
G_{lk} &=& \int\limits_{0}^{1} \left( \prod_{j=1}^{N-1}{\rd}t_j
\; t_j^{a_j-b_j\epsilon}  \right)
\frac{{\cal U}_{lk}^{N_\nu-(L+1)D/2}}{{\cal F}_{lk}^{N_\nu-LD/2}}\, , \quad k=1,\dots ,r\;.
\end{eqnarray}
\end{description}
For each subsector the above steps have to be repeated 
as long as a set ${\cal S}$ 
can be found such that one of the functions 
${\cal U}_{l\dots}$ or ${\cal F}_{l\dots}$ vanishes 
if the elements of ${\cal S}$ are set to zero. 
This way  new subsectors are created in each subsector 
of the previous iteration, resulting in a 
tree-like structure after a certain number of iterations. 
The iteration stops if the functions 
${\cal U}_{l k_1 k_2\dots}$ or ${\cal F}_{l k_1 k_2\dots}$
contain a constant term, i.e.
if they are of the  form
\begin{eqnarray}\label{EQ:subsec_UF}
{\cal U}_{l k_1 k_2\dots} &=& 1 +  u(\vec t\,) \\
{\cal F}_{l k_1 k_2\dots} &=& -s_{0} + 
\sum\limits_{\beta} (-s_{\beta}) f_\beta(\vec t\,)\;, \nonumber
\end{eqnarray}
where $u(\vec t\,)$ and $f_\beta(\vec t\,)$ are polynomials in the
variables $t_j$ (without a constant term), and $s_{\beta}$ 
are kinematic invariants defined by the cuts of the diagram as explained above, 
or (minus) internal masses.
Thus, after a certain number of iterations, 
each integral $G_l$
is split into a certain number, say $\alpha$, of subsector integrals.
We can replace the multi-index $k_1 k_2\dots$ stemming 
from the subsector decomposition by a single index which just counts the 
number of generated subsectors.
The subsector integrals are exactly of the same form as in 
Eq.~(\ref{EQ:subsec_form}), with the difference that the index 
$k$ now runs from 1 to $\alpha$, the total number of produced subsectors
in each primary sector.

Evidently the singular behaviour of the integrand now can be 
read off directly from the exponents $a_j$, $b_j$ for a given subsector integral. 
As the singular behaviour is
manifestly non-overlapping now, it is straightforward
to define subtractions. 

\subsubsection*{Part III \quad Extraction of the poles}

The subtraction of the poles can be done implicitly 
by expanding the singular factors into distributions, 
or explicitly by direct integration over the singular 
factors.
In any case, the following procedure has to be worked through
for each variable $t_{j=1,\dots ,N-1}$ and each
subsector integrand:
\begin{itemize}
\item Let us consider  Eq.~(\ref{EQ:subsec_form}) 
for a particular $t_j$, i.e. let us focus on
\begin{eqnarray}\label{EQsub_step1}
I_j = \int\limits_0^1 dt_j\, t_j^{(a_j-b_j\epsilon)}\, {\cal I}(t_j,\{t_{i\not=j}\},\epsilon) \;,
\end{eqnarray}
where ${\cal I}={\cal U}_{lk}^{N_\nu-(L+1)D/2}/{\cal F}_{lk}^{N_\nu-LD/2}$ 
in a particular subsector.
If  $a_j > -1$, the integration does not lead to an $\epsilon$--pole.
In this case no subtraction is needed and one can go to the next
variable $t_{j+1}$. If  $a_j \leq -1$, one expands ${\cal I}(t_j,\{t_{i\not=j}\},\epsilon)$ into a 
Taylor series around $t_j=0$:
\begin{eqnarray}
{\cal I}(t_j,\{t_{i\not=j}\},\epsilon) &=& \sum\limits_{p=0}^{|a_j|-1}
{\cal I}_j^{(p)}(0,\{t_{i\not=j}\},\epsilon)\frac{t_j^p}{p!} + R(\vec{t},\epsilon) 
\;,\;\mbox{where}\nn\\
{\cal I}_j^{(p)}(0,\{t_{i\not=j}\},\epsilon)&=&
\partial^p {\cal I}(t_j,\{t_{i\not=j}\},\epsilon)/\partial t_j^p\Big|_{t_j=0}\;.
\end{eqnarray}
\item
Now  the pole part can be extracted easily, and one obtains
\begin{eqnarray}\label{EQsub_step2}
I_j = \sum\limits_{p=0}^{|a_j|-1} \frac{1}{a_j+p+1-b_j\epsilon}
 \frac{{\cal I}_j^{(p)}(0,\{t_{i\not=j}\},\epsilon)}{p!} 
+ \int\limits_{0}^{1} dt_j \, t_j^{a_j-b_j \epsilon} R(\vec{t},\epsilon) \;.
\label{tjsubtr}
\end{eqnarray}
By construction, the integral containing the remainder term 
$R(\vec{t},\epsilon)$ does not produce poles in $\epsilon$
upon $t_j$-integration anymore. 
For $a_j=-1$, which is the generic case for renormalisable 
theories (logarithmic divergence), 
this simply amounts to 
$$
I_j=-\frac{1}{b_j\epsilon}\,{\cal I}_j(0,\{t_{i\not=j}\},\epsilon)+
\int\limits_0^1 dt_j\,t_j^{-1-b_j\epsilon}\,
\Big( {\cal I}(t_j,\{t_{i\not=j}\},\epsilon) -{\cal I}_j(0,\{t_{i\not=j}\},\epsilon)\Big)\;,
$$
which is equivalent to applying the 
``plus prescription"~\cite{Gelfand} (see eq.~(\ref{plusdist})), except that
 the integrations over the singular factors have been 
carried out explicitly. 
Since, as long as $j<N-1$, the expression (\ref{tjsubtr}) still contains 
an overall factor $t_{j+1}^{a_{j+1}-\eps\,b_{j+1}}$, it is of the 
same form as (\ref{EQsub_step1}) for $j\to j+1$ and 
the same steps as above can be applied.
\end{itemize}
After $N-1$ steps
all poles are extracted, such that the resulting 
expression can be expanded in $\epsilon$. This defines
a Laurent series in $\epsilon$ with coefficients $C_{lk,m}$ 
for each of the $\alpha(l)$ subsector integrals $G_{lk}$. 
Since each loop can contribute at most one soft and collinear 
$1/\epsilon^2$ term, the highest possible infrared pole of an $L-$loop 
graph  is $1/\epsilon^{2L}$. Expanding to order $\eps^r$, one has
\begin{eqnarray}\label{EQ:eps_series_Glk}
 G_{lk} = \sum\limits_{m=-r}^{2L} \frac{C_{lk,m}}{\epsilon^m} + 
{\cal O}(\epsilon^{r+1}) \;,\quad 
G=(-1)^{N_\nu} \Gamma(N_\nu-LD/2) \sum_{l=1}^{N}\sum_{k=1}^{\alpha(l)} G_{lk}\;.
\end{eqnarray}
Following the steps outlined above one 
has generated a regular integral representation of the
coefficients $C_{lk,m}$, consisting of  $(N-1-m)$--dimensional finite 
integrals over parameters $t_j$.
We recall that  ${\cal F}$ was non-negative
in the Euclidean region where all invariants are negative (see 
eqs.~(\ref{eq0def},\ref{EQ:subsec_UF})),
such that the numerical integrations over the finite parameter 
integrals are straightforward in this region. 
In principle, it is also possible to do at least part of these parameter integrals 
analytically, but in most applications such an analytical approach 
reaches its limits very quickly.

\subsubsection*{Avoiding infinite recursion}

As mentioned already, the choice of the set 
${\cal S}=\{t_{\alpha_1},\dots ,t_{\alpha_r}\}$ which makes 
${\cal U}$ respectively  ${\cal F}$ vanish 
for $t_\alpha\to 0$ is in general 
not unique. The structure of the function ${\cal U}$ 
(see eq.~(\ref{eq0def})) is such that its decomposition will always terminate 
after $L$ iterations for an $L$-loop integral.
For the function ${\cal F}$, the structure depends on the 
masses and kinematic invariants involved. 
Although one could follow one of the mathematical strategies given 
in \cite{Bogner:2007cr} to ensure the iteration terminates, 
this is not the most efficient method for practical purposes, 
as these strategies typically generate a large number of 
subsectors. Another possibility, adopted in~\cite{Denner:2004iz}, is to 
choose the set ${\cal S}$ randomly, such that eventually 
a set will be selected which does not lead to infinite recursion.
However, it is more efficient to use some heuristic rules
which, in all applications to multi-loop diagrams considered 
so far by the author, lead to a terminating decomposition procedure. 

Let us first illustrate the problem by a simple example:
Consider the function 
\be
f(x_1,x_2,x_3)=x_1^2+x_2^2\, x_3\;,
\ee
and suppose we choose ${\cal S}=\{1,3\}$. The replacement 
$x_1=x_3\,t_1$ in the subsector associated with $\theta(x_3-x_1)$ leads to
$\tilde{f}=x_3\,(x_3\,t_1^2+x_2^2)$. Choosing now 
${\cal S}=\{2,3\}$ and substituting $x_2=x_3\,t_2$ in 
the corresponding subsector brings us back to the original 
functional form, so we generate an infinite recursion 
for the above choices of ${\cal S}$. 
In this simple example we can see immediately that the 
choice ${\cal S}=\{1,2\}$ does not lead to this problem.

For multi-loop integrals, we can use the following facts 
as a guideline to choose convenient sets ${\cal S}$:
We first note that an infinite recursion 
does not occur for functions which are linear in each variable. 
The function  ${\cal F}$, before the iterated decompositions, 
is a polynomial 
of maximal degree two in each individual Feynman parameter, where 
quadratic parameters only occur if massive propagators 
are present, due to the term 
${\cal U}(\vec x) \sum\limits_{j=1}^{N} x_j m_j^2$ contained in ${\cal F}$ 
(see eq.(\ref{eq0def})). 
Therefore, a simple extra rule for diagrams with internal masses 
can be added to the procedure:
Before entering the iteration, 
determine the set ${\cal S}_M$ of labels belonging to massive 
propagators and use this set for a first sector decomposition
(even if it does not lead to ${\cal F}=0$ upon setting the elements of ${\cal S}_M$
to zero). This produces a form where in each subsector, 
one of the quadratic powers is reduced by one, such that 
self-similarity to the original form cannot be generated anymore.
In the course of the iterations, quadratic or higher powers will be 
generated unavoidably, such that a form which may lead to 
infinite recursion can occur at some point. In this case it has 
proven useful to choose, if existent, a set ${\cal S}$ containing 
the maximal number of variables occurring with the {\it same} power.
Certainly, these are only heuristic rules, which however 
worked  well in a multitude of practical applications.

\subsection{Examples}

\subsubsection{Planar double box with one off-shell leg}

Two-loop box diagrams with one off-shell leg  are master integrals 
entering for example the two-loop QCD matrix elements for 
 $e^+e^-\to 3$\,jets at NNLO~\cite{Garland:2001tf}. 
Numerical results were first given in \cite{Binoth:2000ps} and served as 
important benchmarks for the analytical calculations of Refs.~\cite{Gehrmann:2000zt,Gehrmann:2001ck}.
\begin{figure}[h]
\begin{center}
\begin{picture}(200,100)
\put(35,5){\includegraphics[width=5.cm, height=2.5cm]{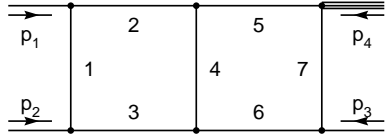}}
\end{picture}
\caption{The planar double-box with leg 4 off-shell.\label{DB1m}}
\end{center}
\end{figure}

For the planar double box with $p_1^2=p_2^2=p_3^2=0, p_4^2\not=0$
shown in Fig.~\ref{DB1m}, the functions ${\cal U}$ and ${\cal F}$ are given by
\begin{eqnarray}
{\cal U} &=& x_{123} x_{567} + x_{4} x_{123567} \nonumber\\
{\cal F} &=& \quad(-s_{12}) (x_2 x_3 x_{4567} + x_5 x_6 x_{1234} 
                         + x_2 x_4 x_6 + x_3 x_4 x_5) \nonumber\\
          &&  +(-s_{23}) x_1 x_4 x_7  
            + (-p_4^2) x_7 ( x_2 x_4 + x_5 x_{1234} ) \;, 
\end{eqnarray} 	    
where $x_{ijk\ldots}=x_i+x_j+x_k+\ldots$ and $s_{ij}=(p_i+p_j)^2$.	        

Iterated sector decomposition produces 197 sectors. 
As the off-shell leg regulates some of the singularities which would be present
in the planar double box with all legs on-shell, the number of 
produced subsectors is lower than for the on-shell 
planar double box (282 subsectors).
The result  
for two Euclidean points is given in Table~\ref{table:db1m}, 
where an overall factor of $\Gamma(1+\eps)^2$ has been extracted
and the integral is defined as\footnote{An infinitesimal imaginary part $+i\delta$ 
in the propagators is understood, and we use $\mu=1$.} 
\bea
DB_{m4}&=&\int \frac{d^D k_1}{i\pi^\frac{D}{2}}
\int \frac{d^D k_2}{i\pi^\frac{D}{2}}\;\times\nn\\
&&\frac{1}{k_1^2\,k_2^2\,(k_1+p_1)^2\,(k_1+p_1+p_2)^2\,(k_2+p_1+p_2)^2\,
(k_2-p_4)^2\,(k_1-k_2)^2}\nn\\
&=&\Gamma(1+\eps)^2\,\left(
\frac{P_4}{\eps^4} +\frac{P_3}{\eps^3} +\frac{P_2}{\eps^2} +
\frac{P_1}{\eps}+ P_0  \right)\;.
\eea
\begin{table}[htb]
\begin{center}
\begin{tabular}{cll} 
\hline
\hline
$(s_{12},s_{23},s_{13},p_4^2)$&$(-1/3,-1/3,-1/3,-1)$ & $(-1/2,-1/3,-1/6,-1)$\\
\hline
\hline
$P_4$&$-26.9997\pm$0.00049&$-11.9998\pm$0.0002\\
$P_3$&$-118.651\pm$0.0037&$-43.0010\pm$0.0027\\
$P_2$&$-239.646\pm$0.0347&$-58.6686\pm$0.0160\\
$P_1$&$-305.823\pm$0.1835&$-20.7692\pm$0.0560\\
$P_0$&$-162.537\pm$0.435&$+98.191\pm$0.289\\
\hline
\end{tabular}
\end{center}
\caption{Numerical results for the pole coefficients of the 
planar double-box with one leg off-shell. An overall prefactor of 
$\Gamma^2(1+\epsilon)$ has been extracted.\label{table:db1m}}
\end{table}
The computing time for the given precision, 
which is better than 0.3\% for the finite part and better than 
0.1\% for the pole coefficients, 
was about 2\,hrs on 3.0\,GHz Intel Xeon processors.
To obtain a precision of only 1\% in the finite part takes about 30 minutes.
The numerical evaluation has been done for each primary sector separately
and the errors have been added in quadrature. 
The independent treatment of each primary sector allows to split the 
problem into smaller subparts which can be evaluated in parallel, 
such that the overall 
computing time is determined by the primary sector with the 
most complicated singularity structure. 
Further, symmetries of the diagram can serve 
as a check, as the results for the corresponding primary sectors 
should be identical. On the other hand, 
if large cancellations  between different primary sectors 
are observed, summing 
over the primary sectors {\it before } the numerical integration 
is the better option.

\subsubsection{Three-loop vertex diagram}

\begin{figure}[htb]
\centerline{\psfig{file=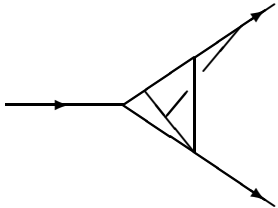,width=3.6cm}}
\caption{A non-planar 3-loop vertex diagram, called $A_8$.\label{a8}}
\end{figure}

As a more complicated example, let us consider the diagram shown in 
Fig.~\ref{a8}, entering the calculation of massless three-loop  
form factors~\cite{Heinrich:2007at}. 
It is given by 
\bea
 A_8 &=& \loopint{k_1}\loopint{k_2}\loopint{k_3}\; \times\\ 
&&\frac{1}{(k_1+p_1)^2 \,  k_2^2 \, 
 k_3^2 \, (k_2+p_1)^2  (k_1+k_3)^2  \,
(k_1+k_3+q)^2 \,(k_2-k_1)^2 \, (k_2+k_3)^2}\nn\;,
\eea
where $q=p_1+p_2$ is the incoming momentum,  and again, 
an infinitesimal imaginary part $+i\delta$ 
in the propagators is understood.
Iterated sector decomposition produced 684 sectors. 
The result is given in Table~\ref{table:A8}, 
where an overall prefactor of 
$i\,S_\Gamma^3\,(-q^2-i\delta)^{-2-3\eps}$ with 
$1/S_\Gamma=(4\pi)^{D/2}\,\Gamma(1-\eps)$, 
 has been extracted:
 \be
A_8 = iS_\Gamma^3\,(-q^2-i\delta)^{-2-3\eps}\,\left(
\frac{P_2}{\eps^2} +\frac{P_1}{\eps}+ P_0+\eps\,P_{\eps}  \right)\;,
 \ee
and the $P_i$ are the coefficients given in Table~\ref{table:A8}.
The computing time up to order $\eps$  
was about 4\,hrs on 3.0\,GHz processors. Computing only 
the pole coefficients and the finite part took about 1 hour.
\begin{table}[htb]
\begin{center}
\begin{tabular}{clr} \hline\hline
&numerical&analytic \\
\hline\hline
$P_2$&$\quad 3.20553\pm 0.00011 $&  3.2054850751\\
$P_1$&$\quad 8.42310\pm 0.00146 $&  8.4222653365 \\
$P_0$& $\quad 27.885\pm 0.039 $& 27.852843117\\
$P_{\eps}$&$-50.246\pm 0.129$&$-50.283167385$\\
\hline
\end{tabular}
\end{center}
\caption{Numerical results for the Laurent-expansion of the 
3-loop vertex diagram shown in Fig.~\ref{a8}. 
The analytic result can be found in~\cite{Heinrich:2007at}.
\label{table:A8}}
\end{table}



\section{Sector decomposition for infrared divergent real radiation integrals}

In order to calculate cross sections at higher orders
in perturbation theory, there are in general not 
only virtual corrections, but also corrections from 
real radiation to be taken into account. At next-to-leading order,
we only have two types of contributions: 
the purely virtual (one-loop) corrections,
and the  real radiation of one additional particle, which 
may be either theoretically or experimentally unresolved.
``Theoretically unresolved" denotes the 
collinear branching of massless particles
or the emission of soft photons or gluons, which
leads to infrared singularities 
appearing as poles in $1/\eps$ in dimensional 
regularisation when integrated over the $D$-dimensional phase space.
Experimentally unresolved particles do not lead to a $1/\eps$-singularity.  
They are defined by 
a so-called ``measurement function" defining the physical observable, 
which in most cases is a subroutine in a numerical program 
rather than an analytic function.
For example, two particles which are clustered into a single jet 
by a certain jet algorithm are considered as experimentally unresolved. 

At NNLO, one generally has to deal with three 
building blocks making up the full cross section: two-loop 
(and one-loop squared) virtual corrections, 
one-loop virtual corrections combined with single unresolved real radiation, 
and doubly unresolved real radiation.

As any $D$-dimensional phase space integral can be transformed to a 
dimensionally regulated multi-parameter integral over the 
unit hypercube, the singularities stemming from the 
(theoretically) unresolved real radiation
are amenable to sector decomposition 
applied to phase space integrals over 
the corresponding squared matrix elements.
What matters here are the {\it denominators} of the  
matrix elements for different processes, which 
have a generic form, and therefore allow for the 
setup of a general framework.

\subsection{Phase space integrals in $D$ dimensions}\label{para}

The phase space integral in $D$ dimensions 
for a generic process $Q\to p_1+\ldots+p_N$ 
can be written as
\begin{eqnarray}
\int d\Phi_{N}^D&=&
(2\pi)^{ N - D (N-1)} \int  \prod\limits_{j=1}^{N} d^Dp_j 
\,\delta^+(p_j^2-m_j^2) 
\delta^{(D)}\Bigl(Q-\sum\limits_{i=1}^{N} p_i \Bigr)\;,
\label{Eq:caseN}
\end{eqnarray}
where 
$\delta^+(p^2-m^2)=\delta(p^2-m^2)\Theta(p^{(0)})$.
Using 
\begin{eqnarray*}
&&\int d^Dp_j\,\delta^+(p_j^2-m_j^2)
=\frac{1}{2E_j}\int
d^{D-1}\vec{p}_j\Big|_{E_j=\sqrt{\vec{p}_j^{\,2}+m_j^2}}\label{dplus}
\end{eqnarray*}
for $j=1,\ldots,N-1$ and eliminating $p_N$ by momentum conservation, 
one obtains
\bea
\int d\Phi_{N}^D&=&\frac{(2\pi)^{ N - D (N-1)}}{2^{N-1}}\int 
\prod\limits_{j=1}^{N-1} 
d^{D-1} \vec{p}_j\,\frac{\Theta(E_j)}{E_j}
 \, \delta^+\Big((Q-\sum\limits_{i=1}^{N-1}
 p_i)^2-m_N^2\Big)\Bigg|_{E_j=\sqrt{\vec{p}_j^{\,2}+m_j^2}}
 \nn\;.
\eea
To proceed, one has to choose a certain parametrisation for the phase space 
integration variables and work out the integration limits confining the 
integration range to the physical region. 
The scattering case $Q=p_a+p_b\to N-1$ particles 
differs from a decay $1\to N$ particles by 
the fact that in the center-of-mass frame of the incident particles, 
it contains a preferred direction
given by the beam axis $\vec{p}_a=-\vec{p}_b$. 
Finding the appropriate phase space integration variables 
which are optimally adapted to 
the kinematic situation at hand can simplify the calculation 
considerably.
This is even more true if sector decomposition is used to 
isolate the infrared singularities: a convenient parametrisation 
will be one where the maximal number of potentially singular 
denominators of the matrix element naturally factorises, thus limiting the number of 
terms produced by iterated decompositions.
In fact, it turns out to be useful to divide the matrix element 
into different ``topologies", according to their denominator structure, 
and use several phase space parametrisations, each being optimal 
for a certain class of topologies. 

\begin{figure}[htb]
\centerline{\psfig{file=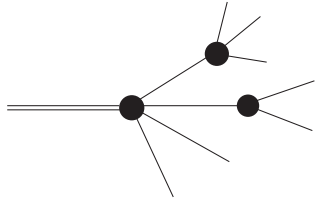,width=4.5cm}}
\caption{Example of a cascade decay corresponding to an iterative 
construction of a multi-particle phase space.\label{cascade}}
\end{figure}
A multi-particle phase space is most conveniently described 
as a convolution of phase spaces of lower multiplicity. 
For example, a process like the one in Fig.~\ref{cascade} 
suggests a phase space parametrisation  which is a convolution 
of a phase space for a $1\to 4$ decay followed by a $1\to 3$ 
decay and a $1\to 2$ splitting.
For a process involving soft radiation off massive fermions, 
it is convenient to choose a parametrisation where the energy of the 
particle which can become soft is an integration variable.
Useful examples of different parametrisations 
can be found e.g. in \cite{Kajantie,Byckling:1970wn,Anastasiou:2005qj}. 
Here, in order to illustrate some generic features of the method, 
we will first derive a phase space parametrisation 
in terms of double invariants $s_{ij}=(p_i+p_j)^2$.

As a pedagogical example we will consider the massless case,  $p_j^2=0$. 
Let use choose a $1\to 4$ process and consider a special frame where 
\bea
Q   &=& (E,\vec 0^{(D-1)})        \nonumber\\
p_1 &=& E_1\, (1,\vec 0^{(D-2)},1) \nonumber\\
p_2 &=& E_2\, (1,\vec 0^{(D-3)},\sin\theta_1,\cos\theta_1)\nonumber\\
p_3 &=& E_3\, (1,\vec 0^{(D-4)},\sin\theta_2\sin\theta_3,\sin\theta_2\cos\theta_3,\cos\theta_2)\nonumber\\
p_4 &=& Q-p_1-p_2-p_3\;,
\eea
which leads to
\bea
d\Phi_{1\to 4} &=& \frac{1}{8}(2\pi)^{4- 3D}\,dE_1\,dE_2\,dE_3\,d\theta_1\,d\theta_2\,d\theta_3
[E_1E_2E_3\sin\theta_1\sin\theta_2]^{D-3}\sin\theta_3^{D-4}\nonumber\\
&&\, d\Omega_{D-2} \; d\Omega_{D-3}\; d\Omega_{D-4}
\Theta(E_1)\,\Theta(E_2)\,\Theta(E_3)\Theta(E-E_1-E_2-E_3)\nonumber\\
&&\delta(E^2-2E(E_1+E_2+E_3)+2(p_1\cdot p_2+p_1\cdot p_3+p_2\cdot
p_3))\;,\label{a11}\\
&&\mbox{where}\nn\\
&&\nonumber\\
\int d\Omega_{D-1}&=&\int_0^{2\pi}d\theta_1\int_0^{\pi}d\theta_2\sin\theta_2\ldots
\int_0^{\pi}d\theta_{D-1}(\sin\theta_{D-1})^{D-2}
=\frac{2\pi^{\frac{D}{2}}}{\Gamma(\frac{D}{2})}\;.\nn
\eea
Now we map the angle and energy variables to 
the double invariants $s_{ij}$ as integration variables, using the Jacobian
\begin{equation}
\det(J) = 
\det\left(\frac{\partial ( s_{..} )}{ \partial (E_i,\theta_j)} \right) =
64\, E^3 E_1^2 E_2^2 E_3^2 \sin\theta_1^2\sin\theta_2^2\sin\theta_3\;.
\end{equation}
The Jacobian in 
combination with terms already present in (\ref{a11}) 
can be written as the determinant $\Delta_4$ of the Gram matrix $G_{ij}=2\,p_i\cdot p_j$.
This determinant  can be expressed by the K\"allen function 
$\lambda(x,y,z)=x^2+y^2+z^2-2xy-2yz-2xz$ as
\bea
\Delta_4 &=& \lambda(  s_{12}\,s_{34}, s_{13}\,s_{24},s_{14}\,s_{23} )=
 -\left( 4\,E\,E_1\,E_2\,E_3\,\sin\theta_1\,\sin\theta_2\,\sin\theta_3 \right)^2\;.
\eea
We see that $\Delta_4$ has to be negative semi-definite.
With the dimensionless variables
\be
y_1 = s_{12}/Q^2\,,\, y_2 = s_{13}/Q^2\,,\,y_3 = s_{23}/Q^2\,,\,y_4 = s_{14}/Q^2\,,\,y_5 = s_{24}/Q^2\,,\,y_6 = s_{34}/Q^2
\label{inv}
\ee
and $\hat\Delta_4 = \lambda(y_1y_6,y_2y_5,y_3y_4)$ we obtain finally 
\bea
d\Phi_{1\to 4} &=& (2\pi)^{4- 3\,D} (Q^2)^{3 D/2-4}\, 2^{-2D+1}\, d\Omega_{D-2} \; d\Omega_{D-3}\; d\Omega_{D-4}\,\nonumber\\&&
\left[ \prod\limits_{j=1}^{6} dy_j \Theta(y_j) \right] \,
\Theta(-\hat\Delta_4)\,[-\hat\Delta_4]^{(D-5)/2}
\delta(1-\sum_{j=1}^{6} y_j)\;.\label{fi4}
\eea
For $N\geq 5$ we have to distinguish 
if we are in $D$ dimensions or in four dimensions. 
In $D$ dimensions, the same procedure as above can in principle be
applied. The four-dimensional case is complicated by the fact that 
the Gram determinant $\Delta_N$ vanishes for $N>4$. 
In this case the phase space can  be expressed in terms of the 
K\"allen function of invariants built from four independent momenta
and additional constraints~\cite{Kajantie}, 
but in practice it is more useful to build it up iteratively as 
described above.

\subsection{Special features of sector decomposition for real radiation}
We see that  expression (\ref{fi4}) has a high symmetry in 
the invariants $y_j$. 
To proceed in a way analogous to the treatment of loop integrals, 
we could now do a ``primary sector decomposition" to 
integrate out the $\delta$-function as 
explained in section \ref{itersd}. 
This would lead to $n$ primary sectors, where $n$ is the number 
of two-particle invariants  $s_{ij}$, i.e. $n=6$ in the example above. 
All invariants are treated on equal footing in this step.
The primary sector decomposition is very useful in the case of 
loop integrals, mainly for the following reason: it preserves the 
feature that singularities {\it only} occur at special {\it points}
at the boundary of parameter space: they occur only if 
$y_{i_1},\ldots ,y_{i_r}= 0$  for a subset $\{i_1\ldots i_r\}$ of 
$\{1\ldots n\}$. 
In other words, 
in the case of loop integrals in the Euclidean region, 
no singularities can occur for $y_i\to 1$ or in the interior of parameter space, 
and by  primary sector decomposition the $\delta$-constraint is integrated out 
without destroying this feature.
In addition, the integration limits  from zero to one for all remaining 
variables are guaranteed without further transformations.

In the case of real radiation, the situation is different, 
because we are forced to stay within the physically allowed region. 
In the parametrisation above, this is reflected 
by the fact that after integrating out the constraint 
$\delta(1-\sum_{i=1}^n y_i)$ 
from momentum conservation, 
we still have the constraint $\Theta(-\Delta_4)$. 
Solving the equation $\Delta_4=0$ 
for say, $y_k$, we  obtain the 
solution 
$y_k^{\pm}=(\sqrt{x}\pm\sqrt{z})^2$, so $y_k^-=0$ whenever $x=z$. 
For example, for $k=3$,  we have 
\be
y_3^{\pm}=(\sqrt{y_2y_5}\pm \sqrt{y_1y_6})^2/y_4\;.
\label{y3pm}
\ee 
The substitution  
$y_3=y_3^-+t_3\,(y_3^+-y_3^-)$  in order to remap all integrations 
to the unit interval will lead to a complicated structure of those denominators 
in the matrix element which contain $y_3$.
In fact, we see that  $1/y_3$ will develop a singularity if  $t_3=0$ and 
$y_3^-$ simultaneously, i.e. 
whenever $t_3=0$ and $y_2y_5=y_1y_6$. 
Thus we found two properties which did not occur in the case of loop integrals:
\begin{enumerate}
\item Square-root terms appear naturally when solving the phase space constraints.
Such terms are potentially dangerous as they may destroy the polynomial 
structure which is a prerequisite for the sector decomposition,   
leading to expressions like 
$g(x,y)=a+y-\sqrt{a^2+x^2}$, where $a$ is a constant. 
However, it is obvious that such terms can be easily transformed into a form 
with the required behaviour under rescaling of the variables. 
\item After having mapped the phase space integration limits to the unit hypercube,
singularities can occur  for a {\em manifold}  which (partly) lies 
{\it inside} the phase space integration region. 
\end{enumerate}
How these singularities can be remapped to the boundaries 
will be shown in a specific example below and 
discussed more generally in Section~\ref{sec:remap}.
Here we would like to point out that we cannot solve the 
constraint $\Theta(-\Delta_4)$ for the same variable $y_k$ in each 
primary sector, because in primary sector $k$, 
$y_k$ has been eliminated. 
Therefore a judicious choice of $y_k$ --- to be an invariant which 
occurs only in very few or no denominators of the 
complete matrix element ---  
would still lead to complicated denominators in  
primary sector $k$, where the constraint 
had to be solved for $y_{i\not=k}$. 
For this reason it is 
advisable not to use primary sector decomposition 
in the case of complicated matrix elements for real radiation. 

\begin{table}[htb]
{\begin{tabular}{lcc} \hline
&loop integrals&phase space integrals\\
\hline
&&\\
parametrisation&${\cal F},{\cal U}$ functions in terms of&many different options, should \\
& Feynman parameters fixed&be adapted to topologies\\
&&\\
primary sector dec.&very convenient&not recommended\\
&&\\
singularity structure&in Euclidean region only &
singularities inside integration \\
&endpoint singularities&region generic\\
&&\\
\hline
\end{tabular}}
\caption{Main differences to loop integrals in the sector 
decomposition procedure for phase space integrals over  IR divergent real radiation
\label{table:diffs}}
\end{table}

To choose a parametrisation which is adapted to the denominator structure 
of the problem, one can follow the idea of iterative splittings outlined above.
Matrix elements involving massless particles contain invariants of the form 
$s_{i_1\ldots i_n}=(p_{i_1}+\ldots +p_{i_n})^2$ with $n\geq 2$ in the denominator. 
\begin{figure}[htb]
\begin{center}
\epsfig{file=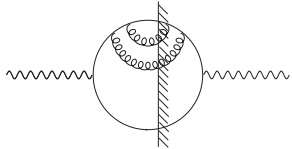,height=2.cm}
\end{center}
\caption{Example of a four-particle cut.\label{Fig:rainbow}}
\end{figure}
For example, the four-particle cut of the diagram in Fig.~\ref{Fig:rainbow} 
contains an integral of the form \cite{Binoth:2004jv}
\bea\label{EqJ4}
J_4 &=& \frac{4}{\pi} \int\limits_{0}^{\infty} \prod_{i=1}^6dy_i\, 
\Theta(-\Delta_4)  (-\Delta_4)^{-1/2-\epsilon} \delta(1-\sum_{j=1}^{6} y_j)
 \frac{(y_1+y_5)\,(y_2+y_6)-y_3\,y_4}{y_2 \,(y_2+y_4+y_6)^2}\;.\nn\\
\eea
In this case, it is suggestive to introduce 
the triple invariant $s_{134}/Q^2=y_2+y_4+y_6$ 
as a genuine phase space variable, such that this denominator 
factorises immediately. This example will be worked out in 
detail in the following section.

Obviously, it is advantageous to 
use  triple invariants as phase space integration variables 
if the amplitude contains a 
splitting of one particle into three final state particles, 
double invariants if the amplitude contains several 
$1\to 2$ splittings, etc. 
Therefore the choice of parametrisation 
is  most conveniently done on a topology basis, i.e. 
different parametrisations are applied to certain classes of denominator 
structures, as already mentioned above.
As the full matrix element contains interferences of 
amplitudes of different ``splitting history", it is in general impossible 
to achieve a factorised form for {\it all} denominators.  
However, minimising the number of decompositions 
by convenient parametrisations is vital to limit the size 
of the expressions produced by iterated sector decomposition. 

The main differences to loop integrals in the sector decomposition procedure
for phase space integrals 
are again summarised in Table~\ref{table:diffs}.

\subsection{Example of a four-particle final state}\label{sec:example}

To explain the concept, we go back to the example 
of the previous section, the 
massless $1\to 4$ phase space, and topologies containing
$s_{134}$  or $s_{234}$  in the denominator.
In order to achieve a convenient parametrisation, we 
first  multiply eq.~(\ref{fi4}) by 
$$1=\int dx_4 \,\delta(x_4-y_2-y_4-y_6)\int d x_5 \,\delta(x_5-y_3-y_5-y_6)$$
and eliminate $y_1,y_4,y_5$ using the $\delta$-functions (see eq.~(\ref{inv}) 
for the definition of the scaled invariants).
Then we solve the constraint 
$(-\Delta_4)\geq 0$
for 
$y_3 = s_{23}/Q^2$ and substitute 
$y_3\rightarrow y_3^-+t_3\,(y_3^+-y_3^-)$.
For the remaining variables we substitute 
\bea
x_4&=&t_4\nn\\
x_5&=&t_1+t_5\,(1-t_1)\,(1-t_4)\nn\\
y_2&=&t_1\,(1-t_2)\,t_4\nn\\
y_6&=&t_1\,t_4\label{remap}
\eea
to arrive at the following form for the phase space
\bea
&&\int d\Phi_{1\to 4} = (2\pi)^{4- 3\,D} (Q^2)^{3 D/2-4}\, 2^{-2D+1}\, d\Omega_{D-2} \; 
d\Omega_{D-3}\; d\Omega_{D-4}\\&&
 \int_0^1 dt_1\ldots dt_5 \,
[(1-t_1)t_4(1-t_4)]^{D-3}[t_1t_2(1-t_2)t_5(1-t_5)]^{\frac{D-4}{2}}
\,[t_3(1-t_3)]^{(D-5)/2}\nonumber
\;.\label{fi4tra}
\eea
The expression for $J_4$ in eq.~(\ref{EqJ4}) then becomes 
\bea
J_4&=&\frac{4}{\pi} \int\limits_{0}^{1} d\,t_1\ldots dt_5\, 
[t_1(1-t_2)]^{-1-\eps}t_4^{-1-2\eps}(1-t_1)^{1-2\eps}(1-t_4)^{2-2\eps}
\nn\\
&&[t_2t_5(1-t_5)]^{-\eps}[t_3(1-t_3)]^{-1/2-\eps}
 \left((t_1(1-t_2)+t_2)-\tilde{y}_3(\vec{t})\right)\\
\tilde{y}_3(\vec{t})&=&t_2t_5+t_1(1-t_2)(1-t_5)-2(1-2t_3)\sqrt{t_1t_2t_5(1-t_2)(1-t_5)}  \nn\\
&=&y_3(\vec{t})/(1-t_4)\nn\;.
\eea
We see that in this parametrisation, 
the denominators in $J_4$ 
are factorising completely. However, other denominators in 
the full matrix element will in general contain $\tilde{y}_3(\vec{t})$ 
in the denominator. 
In this case it is convenient to shuffle the square-root terms 
to the numerator by the following 
non-linear transformation~\cite{Anastasiou:2003gr}: We substitute 
\bea
t_3&=&\frac{y_3^-(1-z_3)}{y_3^-+z_3\,(y_3^+-y_3^-)}\;
\Rightarrow \; y_3=\frac{y_3^+y_3^-}{y_3^-+z_3\,(y_3^+-y_3^-)}\;.
\label{quad}
\eea
The Jacobian therefore cancels one factor of $y_3$ in the denominator:
\be
\frac{{\rm d}t_3}{{\rm d}z_3}=\frac{y_3^+y_3^-}{[y_3^-+z_3\,(y_3^+-y_3^-)]^2}=
\frac{y_3}{y_3^-+z_3\,(y_3^+-y_3^-)}\;,
\ee
leading to 
\bea
&&\int_0^1 {\rm d} t_3 \left[ t_3(1-t_3) \right]^{\frac{D-5}{2}}\,
\frac{1}{y_3(\vec{t})}=\int_0^1 {\rm d} t_3 \left[ t_3(1-t_3) \right]^{\frac{D-5}{2}}\,
\frac{1}{y_3^-+t_3\,(y_3^+-y_3^-)}\nn\\
&=&\int_0^1 {\rm d} z_3 \left[ z_3(1-z_3)\,y_3^+y_3^- \right]^{\frac{D-5}{2}}\,
[y_3^-+z_3\,(y_3^+-y_3^-)]^{4-D}\;.
\eea
This way the square-roots in the denominator are eliminated and the limits 
$t_3\to 0$ and $y_3^-\to 0$ are decoupled, but note that 
instead of $1/y_3(\vec{t})$ we now have a factor 
\bea
\left[y_3^+y_3^- \right]^{\frac{D-5}{2}}=
\left[(1-t_4)\,(t_1(1-t_2)(1-t_5)-t_2t_5) \right]^{D-5}=
(1-t_4)^{D-5}\left[f(t_1,t_2,t_5)\right]^{D-5}\;.\nn
\eea
The factor $(1-t_4)^{D-5}$ will be combined with phase space factors and
is of endpoint-type anyway, but there are singularities which now occur on a manifold 
defined by $f(t_1,t_2,t_5)=0$.
In the case at hand they are easily remapped to the 
boundaries by splitting e.g. the $t_2$-integration region at 
\be
t_2^0=\frac{t_1\,(1-t_5)}{t_5+t_1\,(1-t_5)}
\label{t20}
\ee 
and substituting 
\bea
t_2=t_2^0\,u_2 &&\mbox{ for }  t_2<t_2^0\;,\nn\\ 
t_2=1-(1-t_2^0)\,u_2 &&\mbox{ for } t_2>t_2^0\label{hole}
\eea
to obtain again integrals from zero to one.

\subsection{Possible types of singularities and their treatment}\label{sec:remap}
As we have seen in the previous section, we have to deal with 
the following types of singularities:
\begin{itemize}
\item endpoint singularities
\item singularities on a manifold 
not confined to the boundaries of phase space,  
more precisely the boundaries after having solved all constraints and remapped the integrations to the 
unit hypercube.
\end{itemize}
Endpoint singularities, if not factorising from the start, 
 are easily extracted by the sector decomposition 
algorithm. It should be mentioned however that, 
if we do not use primary sector decomposition, 
endpoint singularities can occur not only if an integration variable 
goes to zero, but also at the upper integration boundary (which is equal to one, after 
appropriate remapping).
In order to apply the algorithm described in 
section \ref{itersd}, we should remap the singularities 
for $t_k\to 1$ such that they occur at the origin only. 
As some variables can cause singularities at zero {\it and } one, 
a transformation $t_k\to 1-t_k$ is not recommended.
Instead, we split the integration range at 1/2:
After the split 
$$\int_0^1 {\rm d} t_k=\underbrace{\int_0^\frac{1}{2} {\rm d} t_k}_{(a)}+
\underbrace{\int_\frac{1}{2}^1 {\rm d} t_k}_{(b)}$$
and the substitution $t_k=u_k/2$ in $(a)$ and $t_k=1-u_k/2$ in $(b)$, 
all endpoint singularities 
occur at $u_k\to 0$ only.
The disadvantage of such splittings is the fact that we end up with  
$2^n$ integrals after $n$ splittings, but in practice, 
considering the physically possible singular limits, some of the 
integration variables clearly never will lead to a singularity 
when approaching one, and therefore do not require such a splitting.

\vspace{3mm}

Concerning the singularities at the interior of phase space, the recipe is 
less simple. However, 
for $N<5$ in a $1\to N$ or a $2\to N-1$ phase space,
it is easy to see that they can always be remapped to the 
phase space boundaries. 
Quite in general, the boundaries of the physical region in the space of 
invariants follow from the momentum conserving $\delta$-function 
and the Gram determinants 
$\Delta_N=\Delta(p_1,\ldots,p_N)=\det(p_i\cdot p_j)$. 
As the Gram matrices are symmetric, the determinants will be polynomials 
of maximal degree two in each invariant $s_{ij}$. Masses do not 
alter this argument, and one can show that always 
 $\Delta_3\geq 0$ and $\Delta_4\leq 0$~\cite{Byckling:1970wn}. 
 The case $N=3$ is trivial, so let us first consider the case $N=4$.
Solving the constraint $\Theta(-\Delta_4)$ 
for one of the invariants $y_k$ 
leads to $y_k^\pm=(\sqrt{a_k}\pm \sqrt{b_k})^2/c_k$, where the structure of $a_k,b_k,c_k$ 
is fixed by the fact that $\Delta_4\leq 0$ is a K\"allen function: 
these terms must be {\it linear} in 
each invariant (see e.g. eq.~(\ref{y3pm})).
After having performed substitutions of the type (\ref{remap}) to eliminate 
the momentum conserving $\delta$-function, the linearity is not 
manifestly preserved, but  as the singularity structure cannot change 
by these substitutions, 
some of the $t_i$ must always factorise, which 
guarantees that  the condition $\Delta_4\leq 0$ 
imposing the phase space boundaries in the new variables  
can be solved for, say, the variable $t_{j}^0$ 
in such a way that $t_{j}^0$ is the ratio of two polynomials in 
the remaining parameters (see e.g. eq.~(\ref{t20})),  
therefore leading to structures amenable to sector decomposition.

\medskip

For  $N\geq 5$, in a $1\to N$ or a $2\to N-1$ phase space, 
additional constraints are present due the fact that, 
for 4-dimensional momenta, 
$\Delta_N=0$ for $N\geq 5$. However, if the phase space is expressed in 
terms of a convolution of processes of lower multiplicity
as explained above, the same reasoning for the remapping of 
singularities as in the $N<5$ case can be applied.

\medskip

Different phase space parametrisations are related by Lorentz 
transformations, therefore it is sufficient to show this property for 
a particular parametrisation. This does not mean that 
all parametrisations actually do have the desired properties, 
it only states that a better parametrisation must exist where the 
remapping to a form more suitable for sector decomposition 
is possible.

\medskip

One last point concerning different types of singularities 
should be made: In renormalisable theories, 
the ``physical" singularities are not worse than logarithmic, 
which means that the parameter integrals 
after sector decomposition should be of the form 
$\int_0^1 dx\,x^{a+b\eps}f(x)$, where $a\geq -1$. 
If only the denominators are considered in a complex matrix element, 
terms with $a<-1$ will occur. This type of  spurious
singularity will finally cancel with terms from the 
numerator, but the cancellation is not manifest if we  
leave the numerator symbolic throughout the whole 
procedure. Therefore it is advisable to include the numerator 
at the level of the $\eps$-expansion, at least for the 
parts where $a<-1$.

\subsection{Construction of a differential Monte Carlo program}

The isolation of infrared poles by sector decomposition is an algebraic 
procedure, leading to a set of finite functions for each pole coefficient 
as well as for the finite part. The finite functions have the form of parameter
integrals over the unit interval and are therefore well suited 
for integration by Monte Carlo methods. 
If a full cross section beyond the leading order, 
composed of both real and virtual corrections, is to be calculated, 
the combination of the sector decomposition approach for the 
real radiation part with analytic results (if available) for the part 
involving loop corrections is certainly possible.
In the case of NNLO corrrections, it is also advisable, 
as the fully numerical evaluation of two-loop integrals in combination with the 
phase space integration is 
in general rather slow, if viable at all. However, at NLO,
examples of a fully numerical evaluation of complete processes 
based on the combination 
of sector decomposition with contour deformation do 
exist~\cite{Lazopoulos:2007ix,Lazopoulos:2007bv}.

A hybrid approach can consist for example in the reduction of the 
phase space integrals to cut master integrals, evaluating 
only the master integrals by sector 
decomposition~\cite{GehrmannDeRidder:2003bm}.
Concerning  
the mixed one-loop times single unresolved real radiation 
part of NNLO calculations, its treatment so far always involved a 
reduction to master integrals~\cite{Anastasiou:2004qd,Anastasiou:2005qj,Anastasiou:2005pn}, 
except in the very recent 
calculation of the $O(\alpha_s^2)$ corrections to semileptonic decay 
$b \to c\, l\,\bar \nu_l$~\cite{Melnikov:2008qs}.
For these mixed real-virtual contributions, 
it can further be useful to do parts of the 
Feynman parameter integrations for the master integrals analytically,
to obtain Hypergeometric functions where 
transformation formulas for the arguments~\cite{Erdelyi,Huber:2005yg} can be 
used if necessary to arrive at a more convenient form.
Then one can proceed by 
applying sector decomposition to the integral representation 
of the Hypergeometric functions in combination with the 
phase space variables~\cite{Heinrich:2004jv,Anastasiou:2005pn}.

To obtain differential results, 
the combination of the output of the 
sector decomposition procedure with any infrared safe 
measurement function is possible, 
as has been first noted in~\cite{Anastasiou:2003gr}. 
The flexibility to do so is achieved by 
expanding the singular factors produced by 
the decomposition into plus-distributions, 
using the identity 
\begin{eqnarray}
\label{plusdist}
&&x^{-1+\kappa\epsilon}=\frac{1}{\kappa\,
\epsilon}\,\delta(x)+
\sum_{n=0}^{\infty}\frac{(\kappa\epsilon)^n}{n!}
\,\left[\frac{\ln^n(x)}{x}\right]_+\;,\nn\\
&&\mbox{where }\nn\\
&&\int_0^1 dx \,f(x)\, \left[g(x)/x\right]_+=\int_0^1 dx \, 
\frac{f(x)-f(0)}{x}\,g(x)\;.
\end{eqnarray}
This is basically equivalent to the $\eps$-expanded form of 
eq.~(\ref{tjsubtr}), the only difference being that, 
instead of integrating out the 
singular factors explicitly, the integrands are 
kept in the form of distributions: 
$\int_0^1 dx\,x^{-1+\kappa\epsilon}\,f(0,y)$ is written as 
$\int_0^1 dx\,\frac{1}{\kappa\,
\epsilon}\,\delta(x)\,f(x,y)$ instead of $\frac{1}{\kappa\,
\epsilon}\,f(0,y)$. This allows for the combination 
with arbitrary functions $f(x,y)$.

The following features which are special to the 
sector decomposition approach as compared to
Monte Carlo programs 
based on analytic subtraction terms should be pointed out:
\begin{itemize}
\item The pole coefficients are only calculated numerically,  
such that the cancellation of poles between 
real, real-virtual (existing beyond NLO only) and purely virtual
contributions can be verified only numerically. 
However, this is in general not a problem because the 
pole coefficients contain less integration variables 
and therefore a high numerical precision 
can be achieved more easily than for the finite part.
\item The expansion into plus distributions 
cannot be done in complete isolation from the
 measurement function, 
because it has to be assured that the subtraction terms only come to 
action in phase space regions which are allowed by the measurement function. 
To illustrate this point, consider the simple one-dimensional example 
where the measurement 
function is just a  step function $\Theta(x-a),\, a>0$, and the 
``matrix element" 
after sector decomposition is given by $f(x)$. 
If we expand into  plus distributions and {\it afterwards} 
just multiply  with our measurement 
function, we obtain
\begin{equation}
\int_0^1\, dx \,\frac{f(x)-f(0)}{x}\,\Theta(x-a)=f(0)\,\ln{a}+
\int_a^1\, dx \,\frac{f(x)}{x}\;.
\label{nosubt}
\end{equation}
Clearly, the 
$f(0)$ term stems from the subtraction of a singularity
at $x=0$, which is now killed by our measurement function anyway, 
such that inclusion of the $f(0)$ term would lead to a wrong result.
The correct way  is 
of course to include the measurement function into the 
expression the plus distribution acts on. 
However, this does {\em not} mean that the $\eps$--expansions and 
subtractions have to be redone each time the measurement function is changed. 
It can be achieved by including symbolic functions in the 
$\eps$--expansion
which are written to the numerical code 
with zero  arguments 
(respectively the appropriate singular limit in the general case) 
if they correspond to 
subtraction terms. The symbolic functions can be specified later
in a subroutine of the numerical program.

\item The  subroutines defining 
jets, observables etc. will be based  on the four-momenta
of the particles involved in the scattering process. 
The four-momenta can easily be constructed from 
the {\it original} phase space integration variables.  
Before the decomposition, the phase space integration variables, 
let us call them $s_{ij}$, have a certain 
functional form, $s_{ij}=s_{ij}(t_1,\ldots,t_n)$. 
Performing now iterated sector decomposition will 
remap the parameters $t_i$, in a different way in each 
subsector of the decomposition tree, such that  
{\it after } iterated sector decomposition, the 
functional dependence of the original 
variables on the Monte Carlo integration parameters 
$t_1,\ldots,t_n$ 
is different for each subsector. 
Of course it is easy to keep track of the remappings 
done in each sector, but the Monte Carlo program 
will consist of a sum of contributions from each subsector $k$, 
each one defining the functional form $s_{ij}^{(k)}(t_1,\ldots,t_n)$
in a different way. 
This is not a problem in principle, 
but the complexity of NNLO matrix 
elements is already enormous, so  
multiplying the evaluation time 
by the number of subsectors, 
which is of the order of several hundreds for an NNLO 
process, can lead to unacceptable CPU times.

\item As the subtractions 
done after sector decomposition are of the form 
$$ \int_0^1 dx \,\frac{f(x)-f(0)}{x}$$ in each variable, 
which means that poles in each variable are {\it locally} 
subtracted, the method in general leads to expressions 
which have a good numerical behaviour. In fact, as even 
integrable singularities of the type $\int_0^1 dx dy\,\frac{1}{x+y}$ 
are decomposed, the expressions produced by iterated sector 
decomposition are of a form which is very convenient 
for numerical integrations. However, 
if the matrix elements to evaluate 
exceed a  certain degree of complexity, there is a turnover 
where the advantage gained from the form of the individual functions 
is destroyed by the sheer number of functions to evaluate. 
This has been found for example in the attempt to calculate 
the full real corrections for $e^+e^-\to 3$\,jets at NNLO
using only sector decomposition.
The calculation of this process has recently been 
accomplished~\cite{GehrmannDe Ridder:2007bj,GehrmannDe Ridder:2007hr,GehrmannDeRidder:2007jk,Ridder:2008ug} 
using analytic ``antenna" 
subtraction~\cite{GehrmannDeRidder:2005cm}. 
Due to the large number of massless particles, 
the infrared structure is extremely complicated, and
the number of antenna subtraction terms needed for the 
 analytic subtraction of the poles is already quite large.
Using sector decomposition leads to 
an unacceptable number of terms in this case. 
On the other hand, if massive particles are involved, 
the situation is completely different:
while analytic integrations of subtraction terms 
become nearly impossible for NNLO calculations with 
several mass scales,  
the infrared singularity structure is less complex in the presence of masses, 
such that the number of terms produced by sector decomposition 
will be moderate, and the mass dependence of the 
finite terms produced by the decomposition does not pose a 
problem for the numerical integration. 
\end{itemize}


\section{Conclusions and outlook}
The method of sector decomposition is interesting 
from a more formal field theoretical point of view as well as 
for phenomenological applications.
Within the context of dimensional regularisation, it offers a 
{\it constructive} scheme for the factorisation and 
subtraction of infrared poles to (in principle) all orders 
in perturbation theory, not only for individual integrals, 
but also for entire squared matrix elements.

Quite in general, the method consists of two parts: 
the first  is an algebraic one, where the singularities are isolated 
in terms of a Laurent series in $\eps$, the coefficients 
being finite parameter integrals. 
The second part consists in the evaluation of these 
parameter integrals, which in general is not possible analytically, 
and is therefore done numerically by  Monte Carlo integration.
Obviously the precision which can be achieved this way is intrinsically 
limited,  compared to the evaluation of functions 
where all integrations have been performed analytically, 
or where deterministic numerical integration methods can be applied.
However, in most practical applications considered so far, 
sufficient precision could be reached within a reasonable 
amount of integration time.

Applications of the method to multi-loop integrals have been very 
successful in providing predictions and 
cross-checks  for cutting-edge analytical 
calculations, e.g. various types of two-loop box integrals 
or three-loop vertex functions. 
A restriction of the method for  multi-loop integrals presented here
is given by the fact the numerical evaluation 
is straightforward only for Euclidean points, where all kinematic 
invariants are negative.  
For one-scale problems, like massless two-point 
or three-point functions, this is not a restriction at all, 
but if more than three external legs or/and masses are 
present, there will be branch cuts and thresholds 
which hinder a straightforward numerical evaluation. 
Solutions to this problem already have been 
suggested~\cite{Binoth:2005ff,Anastasiou:2006hc,Anastasiou:2007qb,Lazopoulos:2007ix,Lazopoulos:2007bv,Melnikov:2008qs,Anastasiou:2008rm}
and are subject to current research.

Although the algorithm is valid to all orders in principle, 
there are certainly  limitations from CPU-time and memory 
once a certain degree of complexity is reached. 
It is not possible to make a general statement about 
where exactly the limit is, as it depends not only on the 
computing resources but also on the way the algorithm 
is programmed. 
Further, the number of loops and scales is not the only 
measure of complexity. Non-planar diagrams in general 
lead to more complicated expressions, often containing
spurious singularities with worse than logarithmic behaviour at 
intermediate stages.

In combination with its application to infrared singular real radiation, 
the method of sector decomposition has proven very useful 
recently to obtain differential results for full processes 
at NNLO~\cite{Melnikov:2008qs,Melnikov:2006kv,Melnikov:2006di,Anastasiou:2005pn,
Anastasiou:2005qj,Anastasiou:2004xq,Anastasiou:2004qd}. 
The main advantages compared to methods based on analytic subtraction 
are the following: the subtraction procedure lends itself to automation
and, due to the ``local" nature of the subtraction terms, 
leads to expressions with good numerical behaviour. 
There is no need for an analytic integration of subtraction terms
over the singular phase space regions, which is a big advantage in the 
presence of massive particles. 
The main drawback of the method consists in the fact that 
it leads to very large expressions for complex processes, 
as the number of terms is increased in each decomposition step. 
In particular for processes involving many massless particles, 
necessitating a large number of subtraction terms, like e.g. 
$e^+e^-\to 3$\,jets or $pp\to 2$\,jets at NNLO,
the size of the expressions produced by sector decomposition 
reaches a limit where  differential results with sufficient 
numerical precision cannot be obtained 
within reasonable CPU times. Fortunately, most processes 
relevant for high precision phenomenology involve both massive 
and massless particles, where the method of sector decomposition
has an enormous potential, not suffering from the 
limitations imposed by analytic integrability.

Part of the problem with intractably large expressions 
is related to the fact that the algorithm, in its fully 
automated form, makes a decomposition already if  the 
{\it necessary} condition to produce a singularity in $1/\eps$  
is fulfilled (e.g. vanishing of the function ${\cal F}$ 
in the case of loop integrals). However, this is 
not always {\it sufficient} to produce a singularity. 
Knowledge about the physical singularity structure
(i.e. the soft and collinear limits) and inspection 
by eye of certain terms can certainly help to 
prevent unnecessary decompositions, but 
the applicability of such criteria is rather limited
for complicated expressions 
as occurring e.g. in NNLO matrix elements, where a fully automated treatment 
is mandatory. Therefore,  
in order to minimise the number of produced terms, it would be
useful to have an  algorithm which finds the 
{\it minimal} number of 
decompositions necessary to extract the singularities.
In the case of scalar loop integrals, this is basically 
a mathematical problem. If full processes are considered, 
a  solution  depends crucially on the way the numerator 
functions are treated. In any case this issue deserves 
further study.

Finally, it is clear that the key to 
an optimal solution often consists in 
combining  several methods in a clever way. The
universality of sector decomposition, as a general method 
to isolate singularities from parameter integrals,  
suggests that it is a good candidate for such combined approaches.

\section*{Acknowledgements}

I would like to thank T.~Binoth 
for fruitful collaboration on 
the subject, especially in what concerns the algorithm for
multi-loop integrals.
I am also very grateful to V.~A.~Smirnov for providing 
analytic challenges of always increasing complexity
allowing to test my program, and for continuous encouragement. 
I also would like to thank A.~Gehrmann-DeRidder and T.~Gehrmann
for useful comparisons and 
encouragement concerning the application to phase space integrals.
Further, I am grateful to T.~Binoth, M.~Czakon, T.~Gehrmann and 
V.~A.~Smirnov for comments on the manuscript.
This research was supported by the UK Science and Technology 
Facilities Council.


\bibliographystyle{JHEP}

\providecommand{\href}[2]{#2}\begingroup\raggedright
\endgroup
\end{document}